\documentclass{optica-article}

\newcommand{\cA}{\hat A}  \newcommand{\ca}{\hat a}
\newcommand{\ba}{\bar a}  \newcommand{\bA}{\bar A}
\newcommand{\dbd}[2]{\frac{\partial #1}{\partial #2}}
\newcommand{\Dint}{D_{\rm int}} \newcommand{\loss}{\frac{l}{2}}
\newcommand{\onum}[1]{{(#1)}}

\renewcommand{\Re}{{\hbox{Re}}}

\journal{opticajournal} 

\articletype{Research Article}
\usepackage[normalem]{ulem}
\usepackage{soul}
\usepackage{multimedia}
\usepackage{hyperref}
\usepackage{media9}
\begin{document}

\title{Multi-color solitons and frequency combs in microresonators}

\author{Curtis R. Menyuk,\authormark{1,*} Pradyoth Shandilya,\authormark{1} 
Logan Courtright,\authormark{1} Gr\'egory Moille,\authormark{2,3} and Kartik Srinivasan \authormark{2,3}}

\address{\authormark{1}Computer Science and Electrical Engineering Department, University of Maryland Baltimore County, Baltimore, MD 21250, USA\\
\authormark{2}Joint Quantum Institute/NIST, University of Maryland, College Park, MD 20742, USA\\
\authormark{3}Microsystems and Nanotechnology Division, National Institute of Standards and Technology, Gaithersburg, MD 20899, USA}

\email{\authormark{*}menyuk@umbc.edu} 


\begin{abstract*} 
Multi-color solitons that are parametrically created in dual-pumped microresonators generate
interleaved frequency combs that can be used to obtain combs at new frequencies and when synchronized can be used
for low-noise microwave generation and potentially as an element in a chip-scale clockwork. Here, we first derive three-wave equations that describe multi-color solitons that appear in microresonators with a nearly quartic dispersion profile.  These solitons are characterized by a single angular group velocity and multiple angular phase velocities.  We then use these equations to explain the interleaved frequency combs that are observed at the output of the microresonator.  Finally, we used these equations to describe the experimentally-observed soliton-OPO effect. In this effect, the pump frequency comb interacts nonlinearly with a signal frequency comb to create an idler frequency comb in a new frequency range, analogous to an optical parametric oscillation (OPO) process. We determine the conditions under which we expect this effect to occur. We anticipate that the three-
wave equations and their extensions will be of use in designing new frequency comb systems and determining their
stability and noise performance.

\end{abstract*}

\section{Introduction}
It was recently discovered that it is possible to synchronously lock two comb teeth over an octave of bandwidth in a single microresonator through an all-optical, $\chi^{(3)}$-mediated process \cite{Moille-2:2023}.  This discovery is a significant step towards the development of a chip-scale optical clock, among other potential applications \cite{Tan-Moss:2023}.  In \cite{Moille-2:2023}, the authors use a primary pump to generate a soliton in the microresonator at around 286 THz and a reference pump at around 194 THz.  The comb lines generated by the soliton are separated by around 1 THz.  When the reference pump is sufficiently close to one of the soliton comb teeth, this tooth is captured by the reference pump.  The back-reaction of the soliton creates a dispersive wave at nearly twice the frequency of the reference pump, which can then be locked by standard $f$-2$f$ self-referencing. {At that point, the entire frequency comb inherits a linewidth on the order of the pump linewidths\cite{Moille-2:2023,Moille-3:2024}}. The comb tooth that is pumped is locked along with all the other comb teeth.

\begin{figure}[ht!]
\centering\includegraphics[width=12cm]{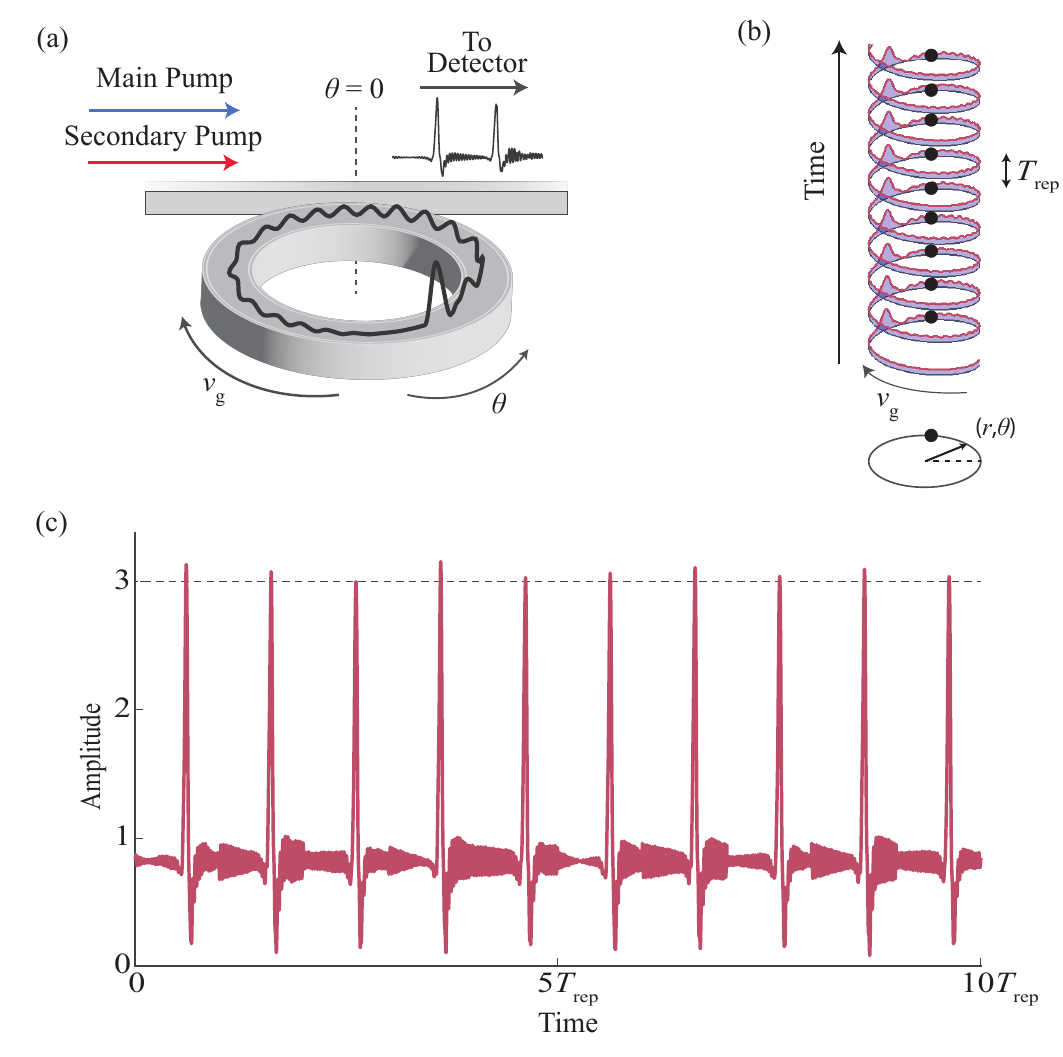}
\caption{Illustration of the stroboscopic process that creates a frequency comb.  In (a), we show the microresonator-waveguide configuration that produces an interleaved frequency comb using both a primary and a secondary pump.  As the soliton, shown propagating in the $-\theta$-direction in (b), circulates around the microresonator, its intensity closest to the waveguide is transferred to the waveguide, which creates a periodic stream of pulses in the output waveguide, as shown in (c).  Visualization 1 shows the time evolution of this process.  When detected at the output of the waveguide, the stream of pulses creates an interleaved comb.}
\label{fig:Fig1}
\end{figure}

This discovery was enabled by the earlier ground-breaking work that established the utility of a secondary pump for taming the thermal instability \cite{Zhang-DelHaye:2019, Lu-Zhao:2019, Zhou-Wong:2019} that had made the generation of soliton frequency combs in microresonators a difficult, non-deterministic process. In \cite{Zhang-DelHaye:2019, Lu-Zhao:2019, Zhou-Wong:2019}, the secondary pump did not participate in the Kerr-mediated processes. Subsequent to that work, it was discovered that by allowing the secondary pump to participate in the Kerr-mediated processes, the secondary pump could also be used to generate {an interleaved comb} with a {total bandwidth of 1.6 octaves in a microresonator with a large free spectral range and a carefully engineered dispersion\cite{Moille:2021}}.  We illustrate this configuration schematically in Fig. \ref{fig:Fig1}(a). It was also observed that the secondary pump produces a secondary frequency comb, referred to here as the signal comb, with the same free spectral range (FSR) as the comb generated by the primary pump, but with a constant frequency offset.  In this experiment, a third set of parametrically-induced comb lines was observed at a frequency on the opposite side of the primary pump frequency, referred to here as the idler comb.  While not reported in \cite{Moille:2021}, this higher-frequency idler comb was observed to have an offset from the primary comb that was just equal and opposite to the offset of the signal comb.

\begin{figure}[ht!]
\centering\includegraphics[width=12cm]{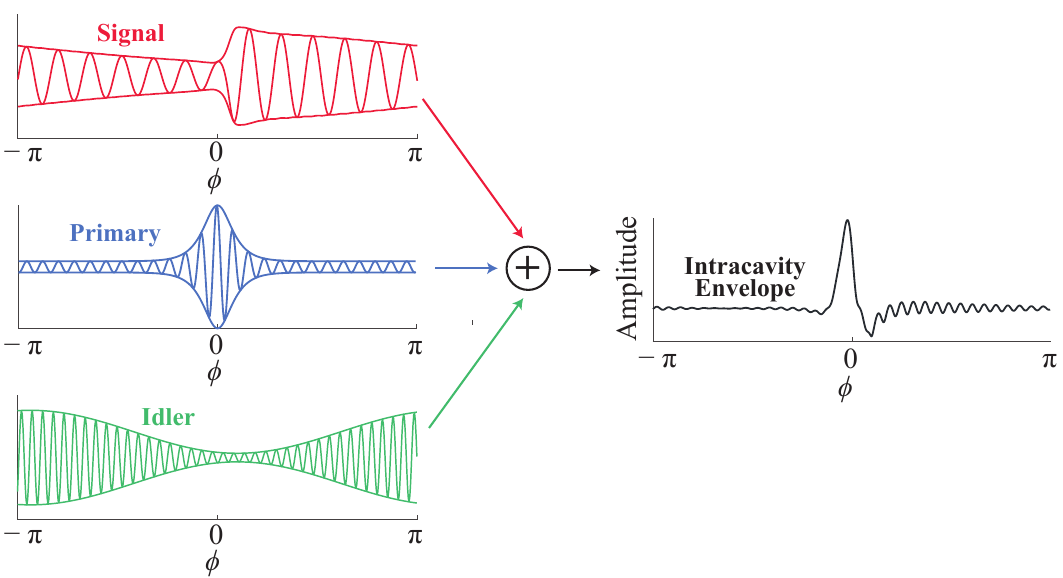}
\caption{A schematic illustration of a multi-color soliton.  Visualization 2 shows the evolution of the waveform as a function of time.  Each component has a well-defined envelope that propagates without changing, as shown on the left. 
 On the right, we show the total field, whose envelope changes as the soliton evolves.}
\label{fig:Fig2}
\end{figure}

A qualitative understanding of this phenomenon was not long in coming.  A multi-color soliton is created in the microresonator cavity.  These multi-color solitons are analogous to solitons at multiple frequencies \cite{Trillo:1988,Yang-Vahala:2017} or two polarizations \cite{Menyuk:1987} that can propagate in optical fibers.  They are also analogous to simultons that have been observed in optical parametric oscillators (OPOs) \cite{Jankowsi-Fejer:2018} and described theoretically in aluminum nitride microcavities \cite{DingYulei-BaoChengying:2024}. While the waveforms in all three colors are nonlinearly locked together and travel around the microresonator with the same (angular) group velocity, their (angular) phase velocities are not the same.  As a result, each color produces a comb with a different carrier envelope offset frequency $f_{\rm ceo}$, and the combs they produce are interleaved.  We schematically illustrate the multi-color soliton that produces interleaved frequency combs in Fig.~\ref{fig:Fig2} and the accompanying visualization. The component of the multi-color soliton that produces the {idler} comb must have $f_{{\rm ceo}+} = 2f_{{\rm ceo}0} - f_{{\rm ceo}-}$ in order to produce the observed frequency offset in the {idler} comb.  In this schematic illustration, the signal comb and idler comb are not localized in the azimuthal angle $\phi$ that circulates around the cavity at the group velocity in contrast to the primary comb, which is localized.  We will show that this behavior is characteristic of the multi-color solitons that generate the interleaved frequency combs. 

By contrast, obtaining a quantitative understanding is not trivial.  It is not sufficient to simply solve a variant of the Lugiato-Lefever equation \cite{Matsko:2011,Coen:2013,Chembo-Menyuk:2013} that has been used with great success to understand many features of microresonator solitons and the combs that they generate.  Within the microresonator, each color has a single frequency in the coordinate system that moves with the soliton's group velocity.  It is only through the detection process when a periodic stream of pulses emerges from the coupling waveguide and is experimentally detected that a frequency comb is created.  To observe the interleaved comb numerically using the Lugiato-Lefever Equation (LLE), one must reproduce this stroboscopic observation process by creating a long stream of pulses and then computing the frequency transform after appropriately filtering the output \cite{Moille-2:2023,Moille-1:2023}. In Fig.~\ref{fig:Fig1} and the accompanying visualization, we show a schematic illustration of this process.

Here, we will show that it is possible to shortcut this numerical procedure and at the same time gain insight into the behavior of these multi-color-soliton waveforms by using three-wave equations that we will derive and that govern the behavior of the solitons that produce the interleaved comb spectrum that was described in \cite{Moille:2021}.  We will show that the higher-frequency {idler} comb previously described is generated through a process that is somewhat analogous to the optical parametric oscillation (OPO) that occurs in continuous wave (CW) systems when a pump wave and a signal wave combine to produce an idler wave.  In contrast however to the case of a CW-OPO, where both the group and phase velocities must be matched, the group velocities are automatically matched via the generation of a multi-color soliton, and it is only necessary to match the phase velocities.  We will then show the precise conditions under which Eqs.~2 and 3 in \cite{Moille:2021} that describe the soliton-OPO effect hold.

The equations that we derive here have applications beyond the explanation of the results in \cite{Moille:2021}.  The integrated dispersion profile in \cite{Moille-2:2023} $D_{\rm int}(\mu)$ is nearly quartic like that of the dispersion profile that produces doubly-dispersive waveguides \cite{Li-Qing-Srinivasan:2017}.  In the case of \cite{Moille:2021}, the zero-crossings are too far away from the primary pump to generate significant dispersive waves.  The spectra in \cite{Moille:2021} and \cite{Li-Qing-Srinivasan:2017} are two examples of a family of spectra that can be produced by similar broadband microresonators.  Other possibilities exist, including the configuration in \cite{Moille-2:2023} in which the reference pump generates a distinct color on the long wavelength side of the primary pump, while a dispersive wave is generated on the short wavelength side.  This entire family of devices can be described using the three-wave equations that we derive here.  {Finally, these equations are the starting point for describing an OPO-soliton cascade like the one observed by Moille et al.~\cite{Moille-1:2023}.}

\section{Derivation of the Three-Wave Equations}
Our starting point is an extended LLE, which in the mode number domain may be written
 \begin{equation} \label{eq:Eq01-LLE-mnd}
    \begin{split}
        \frac{d \cA(m,t)}{dt} &= iD(m)\cA(m) -\frac{l}{2}\cA(m,t) - i\gamma\sum_{n,p}
           \cA(m+n-p,t)\cA^*(n,t)\cA(p,t)\\ 
        & +\sqrt{\kappa_0}\,\delta(m-m_0)E_0\exp(i\omega_0 t)
           +\sqrt{\kappa_-}\,\delta(m-m_-)E_-\exp(i\omega_- t)
    \end{split}
 \end{equation}
where $\cA(m)$ is the field amplitude in the $m$-th azimuthal mode, normalized so that $|A(m)|^2$ is the energy in that mode, $D(m)$ is the linear dispersion of the $m$-th mode, $l$ is the loss, whose $m$-dependence we neglect, $\gamma$ is the Kerr coefficient, whose $m$-dependence we also ignore, $E_0$ and $E_-$ are the primary and secondary pump amplitudes, normalized so that $|E_0|^2$ and $|E_-|^2$ are the pump powers, $m_0$ and $m_-$ are the corresponding mode numbers at which they are resonant, $\sqrt{\kappa_0}$ and $\sqrt{\kappa_-}$ are the corresponding coupling coefficients, and $\omega_0$ and $\omega_-$ are the corresponding angular pump frequencies.  Here, we use $\delta(\cdot)$ to denote the Kr\"onecker delta function or the Dirac delta funcion, depending on whether its argument is discrete or continuous.  In Eq.~\ref{eq:Eq01-LLE-mnd}, where $m$ must be an integer, it denotes a Kr\"onecker delta function.  The sum in Eq.~\ref{eq:Eq01-LLE-mnd} is over all integer values of $n$ and $p$.  In general, discrete sums will denote sums over all possible values.  In the azimuthal domain $(\theta)$, Eq.~\ref{eq:Eq01-LLE-mnd} becomes
 \begin{equation} \label{eq:Eq02-LLE-thd} 
    \begin{split}
        \dbd{\ca(\theta,t)}{t} &= i\sum_m D(m)\cA(m)\exp(im\theta)
        -\frac{l}{2}\ca(\theta,t) - 
            i\gamma|\ca(\theta,t)|^2\ca(\theta,t) \\
        &+\sqrt{\kappa_0}E_0\exp(i\omega_0t + im_0\theta)
            +\sqrt{\kappa_-}E_-\exp(i\omega_0 t + im_-\theta),
    \end{split}
 \end{equation}
where $\ca(\theta,t)$ and $\cA(m,t)$ are an azimuthal Fourier transform pair.  We generally use $x$ and $X$ to denote the Fourier transform pairs that are related via the relations
 \begin{equation} \label{eq:Eq03-FTdef}
   X(m) = \int_{-\pi}^\pi a(\theta)\exp(-im\theta)\,\frac{d\theta}{2\pi},\qquad x(m) = \sum_m X(m)\exp(im\theta).
 \end{equation}
Equation (1) presumes that only one mode family plays a significant role in the dynamics.

We now let $\ca(\theta,t)=\ba(\theta,t)\exp(i\omega_0 t + im_0\theta)$, and we let $\mu = m-m_0$, so that $\mu_- = m_- - m_0$.  Equation (\ref{eq:Eq02-LLE-thd}) then becomes
 \begin{equation} \label{eq:Eq04-LLE-rpv}
    \begin{split}
        \dbd{\ba(\theta,t)}{t} &= i\sum_\mu \left[D(m_0+\mu) - \omega_0\right]\bA(\mu,t)
           \exp(i\mu\theta) -\frac{l}{2}\ba(\theta,t)
           -i\gamma|\ba(\theta,t)|^2\ba(\theta,t) \\
        &\hspace{1cm}+F_0 +F_1\exp\left[i\Delta\omega_- t + i \mu_- \theta\right],
    \end{split}
 \end{equation}
where $\Delta\omega_- = \omega_- - \omega_0$, $F_0 = \sqrt{\kappa_0}\,E_0$, $F_- = \sqrt{\kappa_1}\, E_1$, and $\bA(\mu,t) = \int_{-\pi}^\pi \ba(\theta,t)\exp(i\mu\theta)\,(d\theta/2\pi)$.  We next write $D(m_0 + \mu) = D^{(0)} + D^{(1)}\mu +\Dint(\mu)$, where $D^{(0)}$ and $D^{(1)}$ are the zeroth and first order coefficients of the difference expansion of $D(m_0 + \mu)$.  Equation \ref{eq:Eq04-LLE-rpv} now becomes
 \begin{equation} \label{eq:Eq05-LLE-D-exp}
   \begin{split}
       \dbd{\ba(\theta,t)}{t} &= \left(i\alpha_0-\loss\right)\ba(\theta,t) + D^{(1)}\dbd{\ba(\theta,t)}{\theta} + i\sum_\mu \Dint(\mu)\bA(\mu) \exp(i\mu\theta) \\ 
    &  - i\gamma|\ba(\theta,t)|^2\ba(\theta,t) 
      + F_0 + F_1\exp(i\Delta\omega t +i\mu_-\theta),  
   \end{split}
 \end{equation}
where $\alpha_0 = D^{(0)} - \omega_0 = D(m_0) - \omega_0$ and $\Delta\omega = \omega_- - \omega_0$.

We now specialize to the case where a multi-color soliton has formed so that the waveform is characterized by a single (angular) group velocity, $\omega_{\rm rep} = 2\pi f_{\rm rep} = 2\pi/T_{\rm rep}$, where $f_{\rm rep}$ is the free spectral range (FSR) and $T_{\rm rep}$ is the round-trip time of the soliton.  We can transform the coordinate system in order to freeze the soliton group velocity by letting $\phi = \theta + \omega_{\rm rep} t$, and defining $a(\phi,t) =\ba(\theta+\omega_{\rm rep}t,t)$.  Consistent with Fig.~\ref{fig:Fig1}, we are assuming that the soliton is moving in the $-\theta$ direction.  We then find the multi-pumped Lugiato-Lefever Equation (MLLE)
 \begin{equation} \label{eq:Eq06-LLE-final}
    \begin{split}
        \dbd{a(\phi,t)}{t} & = \left(i\alpha - \loss\right)a(\phi,t) + \beta\dbd{a(\phi,t)}{\phi} + i \sum_\mu \Dint(\mu)A(\mu) \exp(i\mu\phi) \\
       & - i\gamma|a(\phi,t)|^2a(\phi,t) 
      + F_0 + F_1\exp(i\delta\omega t +i\mu_-\phi),
    \end{split}
 \end{equation}
where $\beta = D^\onum{1} - \omega_{\rm rep}$, $A(\mu,t) = \bA(\mu,t)\exp(-i\omega_{\rm rep} t)$, and $\delta\omega = \Delta\omega - \mu_-\omega_{\rm rep}$. In the case of interest to us, the parameter $\beta$ is nearly zero, but it will not be exactly zero since the actual group velocity of the soliton is determined by a weighted mean of the group velocities of both pumps \cite[Supplement 1]{Moille-2:2023}.  The result is a tilt in the experimentally-measured spectrum relative to $D_{\rm int}$ \cite{Moille:2021}.  

In general, Eq.~\ref{eq:Eq06-LLE-final} has no stationary solutions except in the case of Kerr-induced synchronization~\cite{Moille-2:2023}.  The quantity $\delta\omega$ can be as large as $\pm\delta\omega_{\rm rep}/2$.  In the limit of interest to us in which a multi-color soliton has formed, we can obtain a coupled set of equations that has stationary solutions by making the substitution
 \begin{equation} \label{eq:Eq07-Decomp}
    a(\phi,t) = b_-(\phi,t)\exp(i\delta\omega t + i\mu_- \phi)
       + b_0(\phi,t) + b_+(\phi,t)\exp(-i\delta\omega t - i \mu_-\phi),
 \end{equation}
where it is assumed that $b_-$, $b_0$, and $b_+$ all vary slowly in 
time compared to $(\delta\omega)^{-1}$.  We will show that the component proportional to $b_+$ is generated by a four-wave-mixing process that is somewhat analogous to the creation of new frequencies that are generated by continuous wave (CW) OPOs.  In the case of CW-OPOs it is necessary to match both the phase and group velocities.  In the case of a soliton-OPO described here, the group velocities are automatically matched by the Kerr nonlinearity, and it is only necessary to match the phase velocities.  We also note that when the dispersion is sufficiently flat over the bandwidth of the microresonator and $|\mu_-|$ is sufficiently small, it is is possible to cascade the OPO to obtain multiple harmonics, as shown experimentally in \cite{Moille-1:2023}.  This cascade is again somewhat analogous to a similar cascade that can occur  in CW-OPOs.  The cascade can be treated theoretically by letting $a(\phi,t) =\sum_\sigma b_\sigma(\phi,t)\exp(i\sigma\delta\omega t + i\sigma\mu_-\phi)$, where $\sigma$ extends over the range where it is possible to match the phases velocities.  In this paper, we restrict ourselves to the case where $\sigma=\pm1$, which is sufficient to explain the experimental results in \cite{Moille:2021}.

We now substitute Eq.~\ref{eq:Eq07-Decomp} into Eq.~\ref{eq:Eq06-LLE-final}, and we only keep phase-matched terms.  We thus obtain
 \begin{subequations} \label{eq:Eq08-overall}
 {\allowdisplaybreaks
  \begin{align} 
    \begin{split}
        \dbd{b_0(\phi,t)}{t}&=\left(-\loss +i\alpha_0\right)b_0(\phi,t) 
        + i\sum_\mu \left[\beta\mu +\Dint(\mu)\right]B_0(\mu,t)\exp(i\mu\phi) \label{eq:8A-B0}\\
        &\hspace{0.5cm} -i\gamma\left(|b_0(\phi,t)|^2 + 2|b_-(\phi,t)|^2 +2|b_+(\phi,t)|^2\right)b_0(\phi,t)\\ 
        &\hspace{1cm}-2i\gamma b_-(\phi,t)b_+(\phi,t)b_0^*(\phi,t) + F_0,  
    \end{split} \\
    \begin{split} 
       \dbd{b_-(\phi,t)}{t}&=\left[-\loss +i(\alpha_0 - \delta\omega + \beta\mu_-)\right]b_-(\phi,t)\\ &+ i\sum_{\mu'} \left[\beta\mu' +\Dint(\mu'+\mu_-)\right]B_-(\mu',t)\exp(i\mu'\phi) \label{eq:8B-B-}\\
        &\hspace{0.5cm} -i\gamma\left(2|b_0(\phi,t)|^2 + |b_-(\phi,t)|^2 +2|b_+(\phi,t)|^2\right)b_-(\phi,t)\\ 
        &\hspace{1cm} - i\gamma b_0^2(\phi,t)b_+^*(\phi,t) + F_-,
    \end{split} \\
    \begin{split} 
       \dbd{b_+(\phi,t)}{t}&=\left[-\loss +i(\alpha_0 + \delta\omega - \beta\mu_-)\right]b_+(\phi,t)\\ & + i\sum_{\mu''} \left[\beta\mu'' +\Dint(\mu''-\mu_-)\right]B_-(\mu'',t)\exp(i\mu''\phi) \label{eq:8C-B+}\\
        &\hspace{0.5cm}-i\gamma\left(2|b_0(\phi,t)|^2 + 2|b_-(\phi,t)|^2 +|b_+(\phi,t)|^2\right)b_+(\phi,t)\\ 
        &\hspace{1.0cm}-i\gamma b_0^2(\phi,t)b_-^*(\phi,t).
    \end{split}
  \end{align}
  }
\end{subequations}
where $\mu' = \mu-\mu_-$ and $\mu'' = \mu + \mu_-$.  The tilt due the difference between the soliton group velocity and $D^{(1)}$ is now explicit in the terms $i\beta\mu_- b_-$ and $-i\beta\mu_- b_+$ that appear in these equations.  We observe that the $b_+$ component is driven by the term $b_0^2b_-$, which is analogous to the driving term in a CW OPO; the `0' mode is the pump, the `$-$' mode is the signal, and the `+' mode is the idler.  We are assuming that the energy in the 0 mode is large compared to energy in the - mode; otherwise, additional modes could appear in the cascade.  We also observe that typically $|\delta\omega|\gg|\alpha|$ and is on the order of $f_{\rm rep}$.  In order to compensate for this large offset so that the phase velocities match, $\Dint$ must be of the same order.  As a result, the mode numbers at which phase-matching occurs and a resonant interaction is observed are shifted from $\mu_-$ and $\mu_+= -\mu_-$.
    
 \section{Interleaved Frequency Combs and the Soliton-OPO}

\begin{figure}[ht!]
\centering\includegraphics[width=12cm]{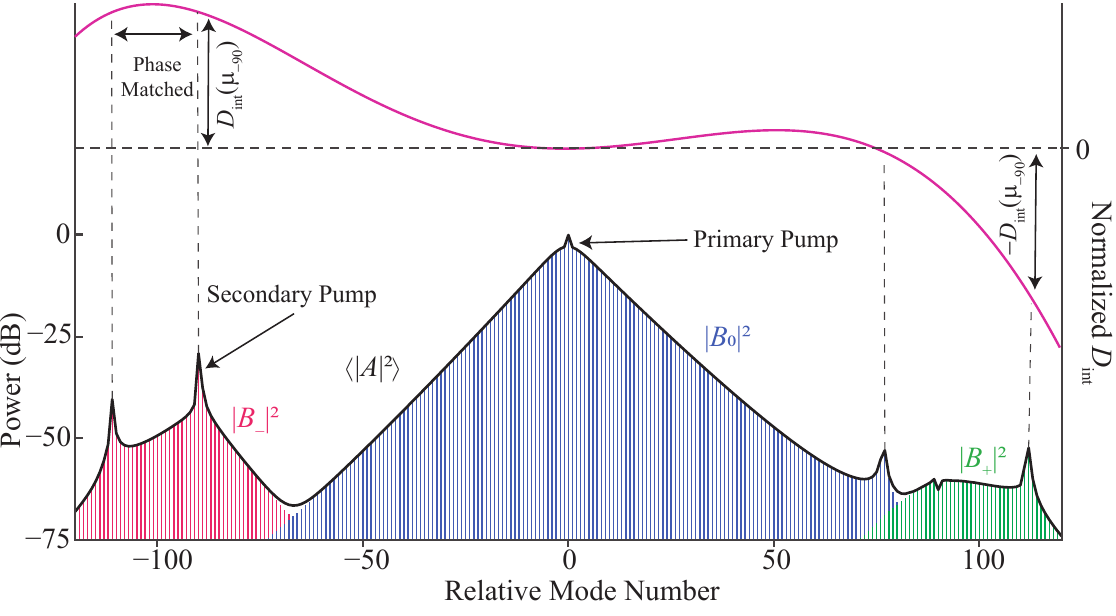}
\caption{Comparison of the average energy $\langle |A|^2\rangle$ of a multi-color soliton that is calculated using Eqs.~\ref{eq:Eq06-LLE-final} and \ref{eq:Eq08-overall} normalized to the primary pump power.  The energy of the three {colors $|B_-|^2, |B_0|^2,$ and $|B_+|^2$} that are computed using {the three-wave equations}, Eq.~\ref{eq:Eq08-overall}, add up to the average energy using {the MLLE}, Eq.~\ref{eq:Eq06-LLE-final}, in the wavenumber domain.}
\label{fig:Fig3-Comp}
\end{figure}

To demonstrate the validity of Eq.~\ref{eq:Eq08-overall}, we compare the solution of Eq.~\ref{eq:Eq08-overall} to Eq.~\ref{eq:Eq06-LLE-final} using the same measured dispersion curve as in \cite{Moille:2021} and that we show in Fig.~\ref{fig:Fig3-Comp}.  We set $F_0=158~\rm{mW}$, $F_-=0.16~\rm{mW}$, and $\mu_-=-90$.  We have $\alpha=-3.3~\rm{GHz}$, $\beta = 2.7~\rm{GHz}$, $\Dint(\mu_-) = 79.7~\rm{GHz}$ and $\Dint(\mu_+) = -19.3~\rm{GHz}$.  The zero-crossings of $\alpha \mp\delta\omega + \beta\mu + \Dint(\mu)$ occur approximately at $\mu=78$ and $\mu=-138$. In Fig.~\ref{fig:Fig3-Comp}, we compare the time average of the energy $|A(\mu,t)|^2$ to the stationary energies $|B_0(\mu,t)|^2$, $B_-(\mu+\mu_,t)|^2$, and $|B_+(\mu-\mu_-,t)|^2$, and the energies in each of the sub-components sums to the energy in $A_0$.

\begin{figure}
\centering\includegraphics[width=12cm]{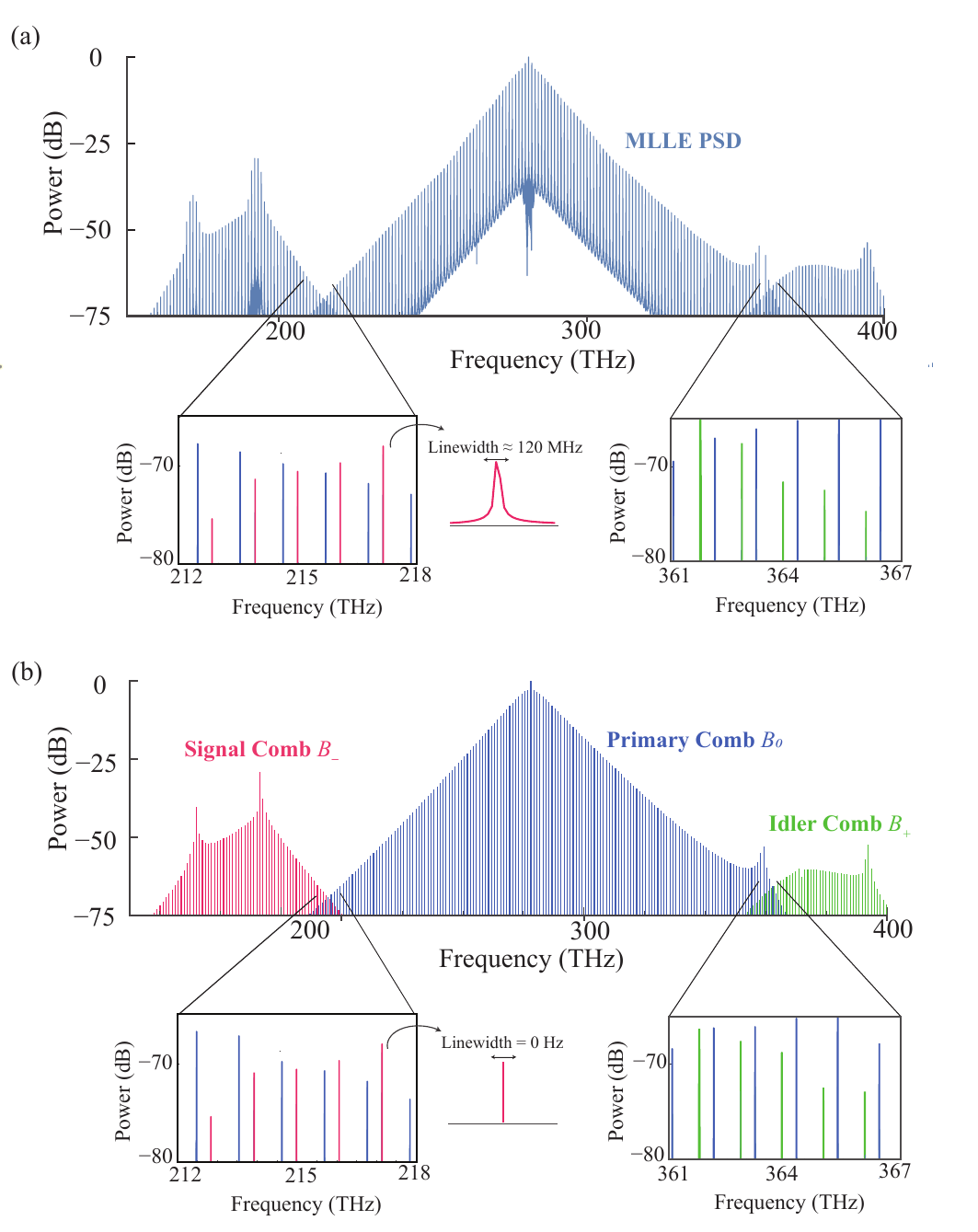}
\caption{The PSD of the output comb, based on (a) Eq.~\ref{eq:Eq06-LLE-final} and on (b) Eq.~\ref{eq:Eq08-overall}.  We show the primary comb $B_0$ in blue, the signal comb $B_-$ in red, and idler comb $B_+$ in green. The PSDs of the calculated combs are nearly identical.  Additionally, we show the lower and upper frequency overlap regions for both calculated combs.  The offset between the lines of the primary comb and idler comb is equal and opposite of the offset between the lines of the primary and signal comb and has a magnitude of 330 GHz.  The results of both calculations are identical.  The only difference is in the linewidth of the individual comb lines, which is approximately 120 MHz in (a) due to the finite number of pulses in the pulse string, but is zero in (b).}
\label{fig:Fig4-PSD}
\end{figure}

It is the observation of a long string of pulses at the output of a coupling waveguide that produces a frequency comb from the soliton that circulates within a microresonator.  In order to observe an interleaved frequency comb when simulating the experimental system \cite{Moille:2021} using Eq.~\ref{eq:Eq06-LLE-final}, we can generate a long string of pulses and compute the string's power spectral density (PSD).  In Fig.~\ref{fig:Fig4-PSD}.a, we show the PSD that is generated using Eq.~\ref{eq:Eq06-LLE-final} for the same systems as in Fig. \ref{fig:Fig3-Comp}.  We used a string of {10~000} pulses and a Hanning filter in order to limit truncation noise.  With {10~000} pulses, the induced linewidth is about 120 MHz, {which is a numerical artifact that occurs due to the finite length of the pulse string.} This linewidth is large compared to the experimentally-observed linewidths\cite{Moille-3:2024}. In Fig.~\ref{fig:Fig4-PSD}.b, we show the PSD that is generated using Eq.~\ref{eq:Eq08-overall}. The PSD in Fig.~\ref{fig:Fig4-PSD}.b is the same as in Fig.~\ref{fig:Fig4-PSD}.a. We show an expanded view of the two overlap regions in both sub-figures.  {We see that the offset between the primary comb $B_0$ and both the signal $B_-$ and the idler $B_+$ has the same magnitude with the opposite sign in both sub-figures.}

The difference between these two combs is visible in the linewidth of the comb lines in the two sub-figures.  The linewidth of the comb lines in Fig.~\ref{fig:Fig4-PSD}.b is zero.  {In the experiments, the actual linewidth when $f$-$2f$ locking is not used is primarily due to a combination of thermorefractive noise and pump noise\cite{Liang-Wong:2017,Stone:2018,Lim-Wong:2019,Lei-Company:2022,Moille-3:2024}.}  These effects could be included in a straightforward way to the comb generated using Eq.~\ref{eq:Eq08-overall}.  Using Eq.~\ref{eq:Eq06-LLE-final}, it would be necessary to use a string that is long enough to resolve the linewidth, which would require on the order of $10^9$ pulses---too long to be practical.

We next turn to the soliton-OPO effect that was observed in \cite{Moille:2021}.  We will assume that the energy in the primary comb is large compared to the energy in the signal, which is again large compared to the energy in the idler comb.  For the simulations in Figs.~\ref{fig:Fig3-Comp} and \ref{fig:Fig4-PSD}, we find $U_0 = \int_{-\infty}^\infty |b_0|^2 d\theta/2\pi = 598\rm{~MJ}$, $U_- = \int_{-\infty}^\infty |b_-|^2 d\theta/2\pi = \rm{9.5~MJ}$, and $U_+= \int_{-\infty}^\infty |b_+|^2 d\theta/2\pi = \rm{6.7~KJ}$; so, this assumption is well-justified.  Given this scaling, we can first linearize Eq.~\ref{eq:8B-B-}, and we obtain
\begin{equation}  \label{eq:Eq09-b--linear}
    \begin{split}
        \dbd{b_-(\phi,t)}{t} &= \left(-\loss + i\alpha_- \right)b_-(\phi,t) + \sum_{\mu'} \left[\beta\mu' + \Dint(\mu'+\mu_-\right]B_-(\mu',t)\exp(-i\mu'\phi) \\
        &\hskip0.5cm -2i\gamma|b_0(\phi,t)|^2b_-(\phi,t) + F_- ,
    \end{split}
\end{equation}
where $\alpha_- = \alpha_0 -\delta\omega + \beta\mu_-$.  In the wavenumber domain, this equation becomes
\begin{equation} \label{eq:Eq10-B--linear}
    \begin{split}
    \frac{d B_-(\mu,t)}{dt} &= \left[-\loss + i\alpha_-  + 
        i\beta\mu' +i\Dint(\mu+\mu')\right]B_-(\mu,t) \\
    &\hspace{0.5cm} - 2i\gamma\sum_{\rho,\sigma}B_0(\rho)B_0^*             (\sigma)B_-(\mu'+\sigma-\rho) + F_-\delta(\mu').
    \end{split}
\end{equation}
We next assume that $B_0(0,t)$ and $B_-(0,t)$ dominate the sum in Eq.~\ref{eq:Eq10-B--linear}.  This assumption is reasonable since they are resonant with the pumps.  The integrated dispersion changes rapidly in the neighborhood of $\mu_-$; so, it is also reasonable to expand $\alpha_- + b\mu' + \Dint(\mu' + \mu_-) \simeq E_- + E_-'\mu'$, keeping only the first two terms in the difference expansion.  Equation \ref{eq:Eq10-B--linear} now simplifies to
\begin{equation} \label{eq:Eq11-B--simplified}
    \begin{split}
        \frac{dB_-(\mu',t)}{dt} &= \left(-\loss + iE_- + iE_-' \mu'\right)B_-(\mu',t) -2i\gamma \left[B_0^*(0,t) B_-(0,t) B_0(\mu',t)\right.\\ 
    &\hspace{0.5cm} \left. + B_0(0,t)B_-(0,t)B_0^*(-\mu',t)\right] 
        + F_-\delta(\mu').
    \end{split}
\end{equation}
In steady state, we then find
 \begin{equation} \label{eq:Eq12-B--SS}
     B_-(\mu') = 2\gamma\left(i\loss + E_- + E_-'\mu'\right)^{-1} 
     \left[B_0^*(0)B_-(0)B_0^*(\mu') + B_0(0)B_-(0)B_0^*(-\mu')\right],
 \end{equation} 
which is resonant when $\mu'\simeq -E_- / E_-'$, as observed in \cite{Moille:2021}.  The mode number spectrum of $B_0$ is imprinted on $B_-$ with an offset that is determined by $\beta$ and $\Dint$ in the neighborhood of $\mu=\mu_-$.

We determine the spectrum of $B_+(\mu'',t)$ in an analogous fashion.  The spectrum of $B_+$ is entirely driven by $B_0$ and $B_-$.  We define $E_+$ and $E_+'$ as the first two terms in the difference expansion of $\alpha +\delta\omega_-$.  We then find in steady state
 \begin{equation} \label{eq:Eq13-B+-SS}
     B_+(\mu'') = 2\gamma\left(i\loss + E_+ + E_+'\mu''\right)^{-1} 
     B_0(0)B_-^*(0)B_0(\mu''),
 \end{equation}
and find that the spectrum of $B_0$ is imprinted on $B_+$ with a resonance at $\mu''\simeq -E_+/E_+'$.  

\begin{figure}[ht!]
\centering\includegraphics[width=12cm]{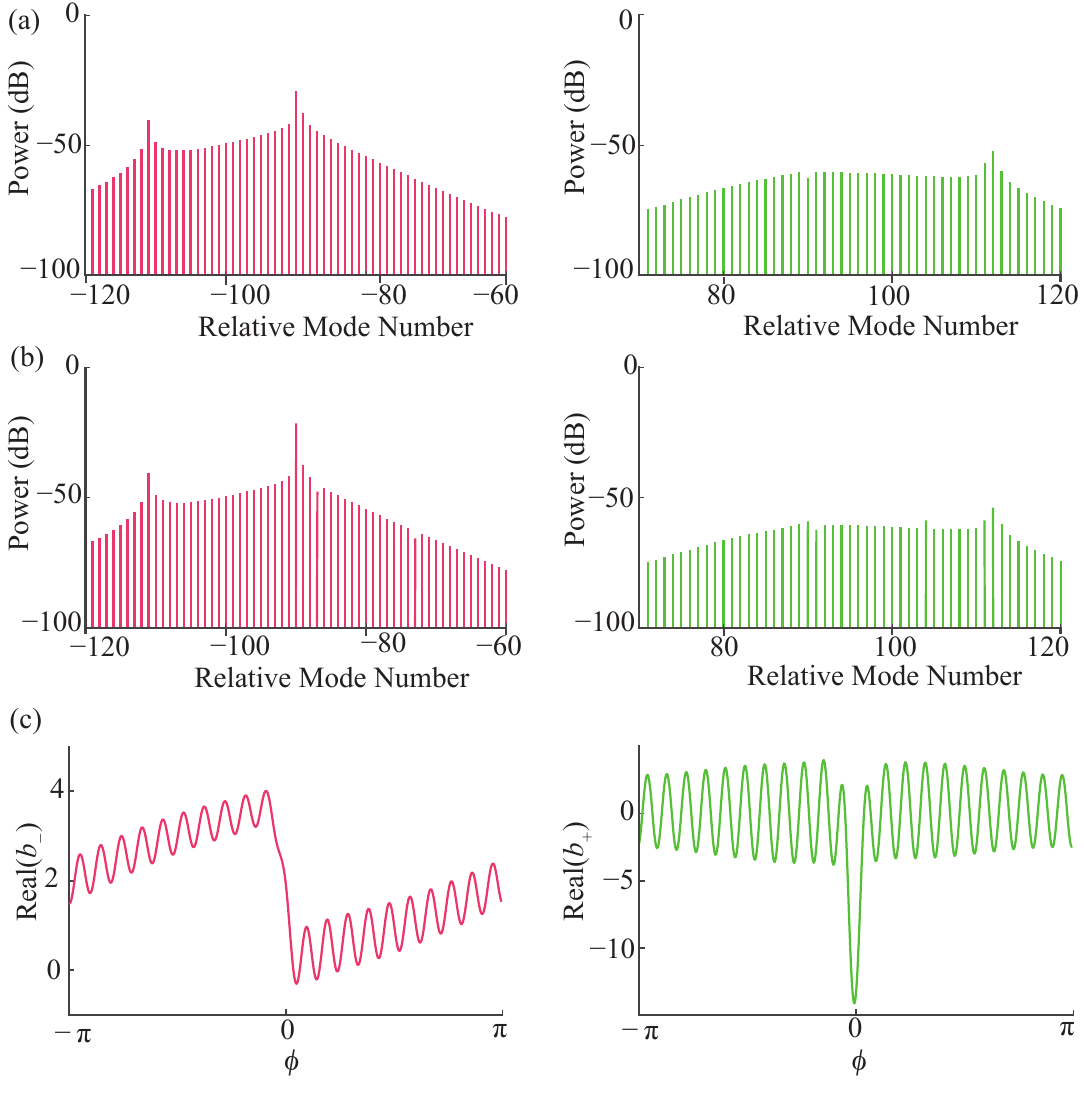}
\caption{We compare $|B_-(\mu')|^2$ and $|B_+(\mu'')|^2$ computed using (a) Eqs.~\ref{eq:8B-B-} and \ref{eq:8C-B+} and using (b) Eqs.~\ref{eq:Eq12-B--SS} and \ref{eq:Eq13-B+-SS}.  In (c), we show $\Re(b_-)$ and $\Re(b_+)$.  The tanh-like shape of $\Re(b_-$) is apparent.}
\label{fig:Fig5}
\end{figure}

In Fig.~\ref{fig:Fig5}, we compare the mode number power spectra that are computed using Eqs.~\ref{eq:8B-B-} and \ref{eq:8C-B+} to the spectra that are computed using Eqs.~\ref{eq:Eq12-B--SS} and \ref{eq:Eq13-B+-SS}.  We see that the results are nearly the same.  In Fig.~\ref{fig:Fig5}.c, we show $\Re(b_-)$ and $\Re(b_+)$.  The tanh-like shape of $\Re(b_-)$ is apparent and will be the subject of a future publication.  This result both explains the soliton-OPO effect that was observed in \cite{Moille:2021} and indicates the conditions that are required to observe it.

\section{Conclusions}

In this paper, we derive three-wave equations that govern the waveform evolution in a microresonator that is pumped at two frequencies in a regime in which a multi-color soliton is produced in the microresonator cavity.  Multi-color solitons are waveforms that have a single group velocity and hence a single $f_{\rm rep}$ while having two or more phase velocities and hence two different carrier envelope offsets $f_{\rm ceo}$.   They appear in microresonators whose integrated dispersion $\Dint$, evaluated at the frequency of the primary pump, is nearly quartic with zero crossings that are far removed from the primary pump. Unless the secondary pump's frequency is tuned close to a comb line of the primary comb, multi-color solitons form. Multi-color solitons play a critical role in the synchronization process.

We began with a derivation of the three-wave equations, and we demonstrated that these equations reproduce the multi-color soliton solutions that are obtained from a modified LLE that uses a realistic dispersion profile.  A periodic stream of multi-color solitons produces an interleaved frequency comb when the stream is detected and its frequency spectrum analyzed outside the microresonator.  Because there are two or more carrier envelope offsets, the relative phase of the colors slip by a constant phase on every round trip.  We showed that the modified LLE can reproduce the observations if one generates a long string of pulses and calculates the Fourier transform of the string.  We then showed that we can use the three-wave equations to shortcut this detection process.

Finally, we used the three-wave equations to explain the soliton-OPO effect that was observed experimentally \cite{Moille-1:2023} and to determine the conditions under which it is expected to appear.  In this microresonator, the zero-crossings of $\Dint$ are too far from the primary pump resonance to generate a doubly-dispersive wave.  However, we show that the primary comb mode number spectrum $B_0$ imprints itself on both the {signal comb} mode number spectrum $B_-$ and the spectrum of the {idler} comb $B_+$ that is generated by the soliton-OPO process.  As a result the total bandwidth of the entire frequency comb can be greatly extended.

There is much that can be done with these equations to study waveforms in microresonator systems.  As one example, they can be used to determine the stability of the muti-color solitons and their response to noise.  These equations can also be used as a starting point for a careful asymptotic analysis of multi-color solitons.  Moreover, the soliton-OPO effect is in many ways analogous to the CW-OPO effect, and it can be cascaded, opening up a path to creating a frequency comb spectrum in otherwise difficult-to-access frequency regimes.  These equations are a useful starting point for analyzing soliton-OPO cascades.  We have barely scratched the surface of what can be learned from these equations.

 \section{Funding}

Work at UMBC was supported by NSF grant ECCS-1807272, by AFOSR grant FA9550-20-1-0357, and by a collaborative agreement 2022138-142232 with the National Center for Manufacturing Sciences as a sub-award from US DoD cooperative agreement HQ0034-20-2-0007.

\bibliography{menyuk}

\begin{thebibliography}{10}
\newcommand{\enquote}[1]{``#1''}

\bibitem{Moille-2:2023}
G.~Moille, J.~Stone, M.~Chojnacky, \emph{et~al.}, \enquote{{K}err-induced synchronization of a cavity soliton to an optical reference,} {\protect\JournalTitle{Nature}} \textbf{624}, 267--274 (2023).

\bibitem{Tan-Moss:2023}
M.~Tan and D.~Moss, \enquote{The trick that could put an optical clock on a chip,} {\protect\JournalTitle{Nature}} \textbf{624}, 256--257 (2023).

\bibitem{Moille-3:2024}
G.~Moille, P.~Shandilya, J.~Stone, \emph{et~al.}, \enquote{All-optical noise quenching of an integrated frequency comb,} arXiv:2405.01238 (2 May 2024).

\bibitem{Zhang-DelHaye:2019}
S.~Zhang, J.~M. Silver, L.~D. Bino, \emph{et~al.}, \enquote{Sub-milliwatt-level microresonator solitons with extended access range using an auxiliary laser,} {\protect\JournalTitle{Optica}} \textbf{6}, 206--212 (2019).

\bibitem{Lu-Zhao:2019}
Z.~Lu, W.~Wang, W.~Zhang, \emph{et~al.}, \enquote{Deterministic generation and switching of dissipative kerr soliton in a thermally controlled micro-resonator,} {\protect\JournalTitle{AIP Advances}} \textbf{9}, 025314 (2019).

\bibitem{Zhou-Wong:2019}
H.~Zhou, Y.~Geng, W.~Cui1, \emph{et~al.}, \enquote{Soliton bursts and deterministic dissipative kerr soliton generation in auxiliary-assisted microcavities,} {\protect\JournalTitle{Light: Science \& Applications}} \textbf{8}, 50 (2019).

\bibitem{Moille:2021}
G.~Moille, E.~F. Perez, J.~R. Stone, \emph{et~al.}, \enquote{Ultra-broadband kerr microcomb through soliton spectral translation,} {\protect\JournalTitle{Nature Communications}} \textbf{12}, 7275 (2021).

\bibitem{Trillo:1988}
S.~Trillo, S.~Wabnitz, E.~M. Wright, and G.~I. Stegeman, \enquote{Optical solitary waves induced by cross-phase modulation,} {\protect\JournalTitle{Optics Letters}} \textbf{13}, 871--873 (1988).

\bibitem{Yang-Vahala:2017}
Q.-F. Yang, X.~Yi, K.~Y. Yang, and K.~Vahala, \enquote{{S}tokes solitons in optical microcavities,} {\protect\JournalTitle{Nature Physics}} \textbf{13}, 53--57 (2016).

\bibitem{Menyuk:1987}
C.~R. Menyuk, \enquote{Stability of solitons in birefringent optical fibers. i: Equal propagation amplitudes,} {\protect\JournalTitle{Optics Letters}} \textbf{12}, 614--616 (1987).

\bibitem{Jankowsi-Fejer:2018}
M.~Jankowski, A.~Marandi, C.~R. Phillips, \emph{et~al.}, \enquote{Temporal simultons in optical parametric oscillators,} {\protect\JournalTitle{Physical Review Letters}} \textbf{120}, 053904 (2018).

\bibitem{DingYulei-BaoChengying:2024}
Y.~Ding, Z.~Wei, Y.~Wang, \emph{et~al.}, \enquote{Theoretical analysis of microavity simultons reninforced by $\chi^{(2)}$ and $\chi^{(3)}$ nonlinearities,} {\protect\JournalTitle{Physical Review Letters}} \textbf{132}, 013801 (2024).

\bibitem{Matsko:2011}
A.~B. Matsko, A.~A. Savchenko, W.~Liang, \emph{et~al.}, \enquote{Mode-locked frequency combs,} {\protect\JournalTitle{Optics Letters}} \textbf{36}, 2845--2847 (2011).

\bibitem{Coen:2013}
S.~Coen, H.~G. Randle, T.~Sylvestre, and M.~Erkintalo, \enquote{Modeling of octave-spanning {K}err frequency combs using a generalized mean-field {L}ugiato–{L}efever model,} {\protect\JournalTitle{Optics Letters}} \textbf{38}, 37--39 (2013).

\bibitem{Chembo-Menyuk:2013}
Y.~K. Chembo and C.~R. Menyuk, \enquote{Spatiotemporal lugiato-lefever formalism for {K}err-comb generation in whispering-gallery-mode resonators,} {\protect\JournalTitle{Physical Review A}} \textbf{87}, 053852 (2013).

\bibitem{Moille-1:2023}
G.~Moille, C.~Li, J.~Stone, \emph{et~al.}, \enquote{Two-dimensional nonlinear mixing between a dissipative {K}err soliton and continuous waves for a higher-dimension frequency comb,} arXiv:2303.10026v2 (20 Mar 2023).

\bibitem{Li-Qing-Srinivasan:2017}
Q.~Li, T.~C. Briles, D.~A. Westly, \emph{et~al.}, \enquote{Stably accessing octave-spanning microresonator frequency combs in the soliton regime,} {\protect\JournalTitle{Optica}} \textbf{4}, 193--203 (2017).

\bibitem{Liang-Wong:2017}
J.~Lim, A.~A. Savchenkov, E.~Dale, \emph{et~al.}, \enquote{Chasing the thermodynamical noise limit in whispering-gallery-mode resonators for ultrastable laser frequency stabilization,} {\protect\JournalTitle{Nature Communications}} \textbf{8}, 8 (2017).

\bibitem{Stone:2018}
J.~R. Stone, T.~C. Briles, T.~E. Drake, \emph{et~al.}, \enquote{Thermal and nonlinear dissipative-soliton dynamics in {K}err-microresonator frequency combs,} {\protect\JournalTitle{Physical Review Letters}} \textbf{12}, 063902 (2018).

\bibitem{Lim-Wong:2019}
J.~Lim, W.~Liang, A.~A. Savchenkov, \emph{et~al.}, \enquote{Probing 10 \textmu k stability and residual drifts in the cross-polarized dual-mode stabilization of single-crystal ultrahigh-q optical resonators,} {\protect\JournalTitle{Light: Science and Applications}} \textbf{8}, 1 (2019).

\bibitem{Lei-Company:2022}
F.~Lei, Z.~Ye, O.~B. Helgason, \emph{et~al.}, \enquote{Optical linewidth of soliton microcombs,} {\protect\JournalTitle{Nature Communications}} \textbf{13}, 3161 (2022).

\end{thebibliography}

\end{document}